\documentstyle[12pt,epsf,epsfig,amsmath,amsfonts,color]{article}

\def\be{\begin{equation}}
\def\ee{\end{equation}}
\def\ba{\begin{eqnarray}}
\def\ea{\end{eqnarray}}

\def\bdm{\begin{displaymath}}
\def\edm{\end{displaymath}}
\def\la{~\mbox{\raisebox{-.6ex}{$\stackrel{<}{\sim}$}}~}

\def\bq{\begin{quote}}
\def\eq{\end{quote}}

 at 10truept

\newcommand{\beq}{\begin{equation}}
\newcommand{\eeq}{\end{equation}}
\newcommand{\bea}{\begin{eqnarray}}
\newcommand{\eea}{\end{eqnarray}}
\newcommand{\beqa}{\begin{eqnarray}}
\newcommand{\eeqa}{\end{eqnarray}}

\def\la{~\mbox{\raisebox{-.6ex}{$\stackrel{<}{\sim}$}}~}

\def\ltap{\ \raise.3ex\hbox{$<$\kern-.75em\lower1ex\hbox{$\sim$}}\ }
\def\gtap{\ \raise.3ex\hbox{$>$\kern-.75em\lower1ex\hbox{$\sim$}}\ }
\def\gl{\ \raise.5ex\hbox{$>$}\kern-.8em\lower.5ex\hbox{$<$}\ }
\def\roughly#1{\raise.3ex\hbox{$#1$\kern-.75em\lower1ex\hbox{$\sim$}}}

\begin{document}

\thispagestyle{empty}
\begin{flushright}
May 2015
\end{flushright}
\vspace*{1cm}
\begin{center}
{\Large \bf A Manifestly Local Theory of Vacuum Energy}
\vskip.3cm
{\Large \bf Sequestering}

\vspace*{1cm} {\large Nemanja Kaloper$^{a, }$\footnote{\tt
kaloper@physics.ucdavis.edu}, Antonio Padilla$^{b, }$\footnote{\tt
antonio.padilla@nottingham.ac.uk}, David Stefanyszyn$^{b, }$\footnote{\tt
ppxds1@nottingham.ac.uk}}\\
\vskip.3cm
{\large and George Zahariade$^{a, }$\footnote{\tt
zahariad@ucdavis.edu}}\\
\vspace{.5cm} {\em $^a$Department of Physics, University of
California, Davis, CA 95616, USA}\\
\vspace{.3cm} {\em $^b$School of Physics and Astronomy, 
University of Nottingham, Nottingham NG7 2RD, UK}\\

\vspace{1.5cm} ABSTRACT
\end{center}
We present a manifestly local, diffeomorphism invariant and locally Poincare invariant formulation of vacuum energy sequestering. In this theory, quantum vacuum energy generated by matter loops is  cancelled by auxiliary fields. The auxiliary fields  decouple from gravity almost completely. Their only residual effect is an
{\it a priori} arbitrary, finite contribution to the curvature of the background geometry, which is radiatively stable. Its value is to be determined by a measurement, like the finite part of any radiatively stable UV-sensitive quantity in quantum field theory.

\vfill \setcounter{page}{0} \setcounter{footnote}{0}
\newpage

In a recent series of papers \cite{KP1, KP2, KP3} two of the authors have suggested the mechanism for sequestering vacuum energy. The idea was to 
gravitationally decouple vacuum energy generated by matter loops\footnote{The mechanism ignores the graviton loops. However the matter loops alone render the cosmological constant radiatively unstable unless the matter is conformal and/or supersymmetric; see
\cite{zeldovich,wilczek,wein,pol,me}.}, providing a workaround for Weinberg's no-go theorem obstructing dynamical adjustment \cite{wein}. 
The mechanism is based on two new rigid variables without {\it any} local degrees of freedom: the bare cosmological constant $\Lambda$ and the scaling parameter $\lambda$ measuring the matter sector dimensional scales in Planck units. They are Lagrange multipliers, enforcing two global constraints, in a way similar to the isoperimetric problem of variational calculus \cite{elsgolts}.  The action principle with these fields rests on  \cite{KP1, KP2}
\be
S= \int d^4 x \sqrt{g} \left[ \frac{M^2_{Pl}}{2} R  - \Lambda - {\lambda^4} {\cal L}_m(\lambda^{-2} g^{\mu\nu} , \Phi) \right] +\sigma\left(\frac{ \Lambda}{\lambda^4 \mu^4}\right) 
 \, , ~~
\label{action}
\ee
where the global interaction term $\sigma\left(\frac{ \Lambda}{\lambda^4 \mu^4}\right) $  is {\it outside} of the integral. The function $\sigma$ is required to be an odd differentiable function, and the mass scale $\mu$ is around the QFT cut-off.  $\Phi$ correspond to the `protected' matter fields, e.g., the Standard Model.

The variation of (\ref{action}) with respect to $\Lambda$ links the gauge invariant global 4-volume
$\int d^4x \sqrt{g}$ to $\lambda$, making all matter sector scales dependent on $\int d^4x \sqrt{g}$. In turn, the variation of 
(\ref{action}) with respect to $\lambda$ yields a dynamical constraint $\Lambda =  \langle T^\alpha{}_\alpha \rangle/4$, where $T^\mu{}_\nu$ is the energy momentum tensor of the canonically normalized matter fields, and $\langle Q\rangle ={\int d^4 x \sqrt{g} \, Q}/{\int d^4 x \sqrt{g}}$ is the world-volume average of any $Q$. Thus the gravitational field equations reduce to
\be
M_{Pl}^2 G^\mu{}_\nu= T^\mu{}_\nu-\frac{1}{4} \delta^\mu{}_\nu \langle T^\alpha{}_\alpha \rangle \, .
\label{eeqs}
\ee
The matter sector quantum corrections to vacuum energy all scale as $\lambda^4$ \cite{selft}. They are all accounted for in 
$\langle T^\alpha{}_\alpha \rangle$, and cancel automatically from the right hand side of (\ref{eeqs}). Extracting the constant contribution to the stress energy, $V_{vac}$, and rewriting stress energy as   $T^{\mu}{}_\nu = -V_{vac} \, \delta^\mu{}_\nu + \tau^\mu{}_\nu$, where $\tau^\mu{}_\nu$ describe local excitations, we see that $V_{vac}$ completely drops out from (\ref{eeqs}). There remains a residual  cosmological constant: the historic average  $\langle \tau^\mu{}_\mu \rangle/4$ which is completely insensitive to vacuum loop corrections, and is small in large and old universes \cite{KP1,KP2}\footnote{This is thanks to two approximate symmetries of the theory, broken only by the 
gravitational sector: the scaling $\lambda \rightarrow \Omega \lambda$, $g_{\mu\nu} \rightarrow \Omega^{-2} g_{\mu\nu}$ and $\Lambda \rightarrow \Omega^4 \Lambda$, and the
shift $\Lambda \rightarrow \Lambda + \alpha \lambda^4$ and ${\cal L}_m \rightarrow {\cal L}_m -\alpha$. They ensure that the vacuum energy cancels independently of the scale, and make a small residual cosmological constant natural since they are enhanced in the conformal limit describing infinite conformal universes.}. It must be nonlocal, since it is the renormalized, finite part of the cosmological constant: so it must be {\it measured}, as any leftover of a UV-sensitive quantity in QFT. Since the cosmological constant is a spacetime filling quantity the only detector which can measure it with arbitrary precision is the whole universe, implying a nonlocal measurement \cite{degrav,KP1,KP2}. Further the original mechanism requires finite spacetime volume to accommodate nonzero matter sector mass scales.

Even so, the global term $\sigma\left(\frac{ \Lambda}{\lambda^4 \mu^4}\right)$ is unusual. Although the `on-shell nonlocality' it induces by fixing the residual cosmological constant is harmless - and indeed necessary - in QFT coupled to (semi) classical gravity,
this term appears to conflict with the expectations about the microscopic origin of the mechanism. If the action (\ref{action}) is 
a low energy limit of some theory of quantum gravity, one expects that the underlying theory has a 
Feynman's path integral \cite{feyn}. Yet $\sigma\left(\frac{ \Lambda}{\lambda^4 \mu^4}\right)$ seems to obstruct this because it does not appear to be additive over spacetime. Terms like it arise as low energy corrections to the action from quantum gravity effects, described by wormhole calculus  \cite{cole}. They have been argued to appear if there are locally separated, but quantum-entangled duplicate universes \cite{andreimult}, where one copy could even be compactified \cite{tseytlin}. Here we will seek a simpler 
route and show that global terms like $\sigma\left(\frac{ \Lambda}{\lambda^4 \mu^4}\right)$ can be thought of as conserved quantities in a {\it manifestly local} theory. We will also show that the rigid variables $\Lambda,\lambda$ are solutions of local field equations which only admit constant roots thanks to the gauge redundancies of the extra sectors. The resulting theory sequesters the matter-generated quantum vacuum energy in almost exactly the same way as the original proposal \cite{KP1,KP2,KP3}.
The main differences are that the residual cosmological constant is not uniquely fixed in terms of other matter sources, but involves an arbitrary integration constant (actually, a ratio of two constants), and that the spacetime volume of the underlying geometry does not have to be finite. 
This is the `price to pay' (or a `reward to reap') when  interpreting the two global constraints as solutions of two local field equations. However, the residual cosmological constant -- a finite part of a UV-sensitive quantity -- must be measured rather than computed. Since it is radiatively stable, its value can now be evaluated reliably. 

Let us now turn to the construction of a local theory which sequesters matter sector vacuum energy. For reasons of calculational simplicity, 
we can absorb $\lambda$ into the definition of the Planck scale, going from `Einstein frame'  to `Jordan frame' variables by 
$g_{\mu\nu} \to \frac{\kappa^2}{M_{Pl}^2} g_{\mu\nu}$, $\Lambda \to \Lambda \left(\frac{M_{Pl}^2}{\kappa^2}\right)^2$
where we have defined the new variable $\kappa^2 = {M_{Pl}^2}/{ \lambda^2 }$. 
The action now reads
\be
S= \int d^4 x \sqrt{g} \left[ \frac{\kappa^2}{2} R  - \Lambda - {\cal L}_m( g^{\mu\nu} , \Phi) \right] +\sigma\left(\frac{ \Lambda}{ \mu^4}\right) 
 \, . ~~
\label{actionJ}
\ee
Varying with respect to the rigid variable $\kappa^2$ yields a
global constraint $\langle R \rangle =0$. The metric variation gives the standard gravitational field equations. Combining them with $\langle R \rangle =0$ yields $\left(\Lambda-\frac14 \langle T^\alpha{}_\alpha \rangle \right)/{\kappa^2}=0$. When we do not decouple gravity, so $\kappa^2$ is finite, this shows that $\Lambda = \langle T^\alpha{}_\alpha \rangle/4$, as before. 
Further, $\Lambda$ variation gives $\int d^4 x \sqrt{g}=\sigma'/\mu^4$. If $\int d^4 x \sqrt{g}$ diverged, $\Lambda$ would have to be a singular point of  $\sigma'$, and moreover the constraint $\frac{1}{\kappa^2}\left(\Lambda-\frac14 \langle T^\alpha{}_\alpha \rangle \right)=0$ could only be satisfied 
with $\kappa^2$ infinite. The ratio of the particle masses to the Planck scale $\kappa^2$ would vanish then, forcing the matter sector to be massless. 
Thus as in the `Einstein frame', the worldvolume of the universe must be finite to accommodate nonzero matter sector mass scales. Note that 
$\kappa^2$ and $\Lambda$ are taken to be rigid quantities,
without any local degrees of freedom, being uniform throughout spacetime, even though we are to vary (\ref{action}) with respect to both of them. 
Our aim is to relax this, and change the action (\ref{actionJ}) so that
it is completely local, and yet leads to qualitatively the same field equations.

A clue how to do this is provided in the discussion of the isoperimetric problem in the calculus of variations  in \cite{elsgolts}. The idea is to 
interpret the global constraint which fixes the perimeter of a curve as an integral of its first derivative, and enforce a constraint on it by way of a local Lagrange multiplier. Since the local constraint involves only a first derivative, the solution for the Lagrange multiplier must be a constant, but now
this is a consequence of the equations of motion rather than an external assumption.
The cost is to allow 
the perimeter to be an arbitrary integration constant rather than a fixed number. Nevertheless, the variational procedure then picks an extremal
surface for each given value of the perimeter. The action is additive, and one can define the Hamiltonian and
the path integral for it.

We will adopt this procedure to the case of a QFT coupled to gravity, starting with (\ref{actionJ}). We wish to promote the rigid parameters $\kappa^2, \Lambda$ to local fields, and reinterpret the global term as an integral of local expressions, which simultaneously yield 
local equations  $\partial_\mu \kappa^2 = \partial_\mu \Lambda = 0$. The new local additions should not gravitate directly in order to preserve the main feature of sequestering: the matter-induced quantum 
vacuum energy needs to completely drop out. The route to follow has already been hinted at in the gauge invariant formulation of unimodular quantum gravity by Henneaux and Teitelboim in 1989 \cite{HT}. To enforce the constraint
$\sqrt{g} = 1$ in a way which manifestly respects diffeomorphism invariance, instead of adding $\int d^4x \, \Lambda(x) \, (\sqrt{g} - 1)$ to the Einstein-Hilbert action, and treating $\Lambda(x)$ as a Lagrange mutliplier,
one adds a term with a {\it different measure}. Instead of $\sqrt{g}$, any determinant works in its stead. Since 
the covariant measure is $dx^\mu dx^\nu dx^\lambda dx^\sigma \sqrt{g} \epsilon_{\mu\nu\lambda\sigma}/4!$, one 
can replace $ \sqrt{g} \epsilon_{\mu\nu\lambda\sigma}$ by {\it any} 4-form $F_{\mu\nu\lambda\sigma}$, using the gauge fixing term
\be
S_{GF} = - \int  \Lambda(x) \, \left(\sqrt{g} \, d^4x - \frac{1}{4!} F_{\mu\nu\lambda\sigma} \, dx^\mu dx^\nu dx^\lambda dx^\sigma\right) \, .
\label{gfterm}
\ee
Since $F$ is completely independent of the metric off shell, it does not appear in the gravitational field equations at all.  
If we define $F_{\mu\nu\lambda\sigma} = 4 \partial_{[\mu} A_{\nu\lambda\sigma]}$, where the square brackets denote antisymmetrization of enclosed indices, then the field equation obtained by varying $ A_{\nu\lambda\sigma}$ is just $\partial_\mu \Lambda(x) = 0$, fixing the
Lagrange multplier $\Lambda(x)$ to be an arbitrary constant -- i.e., rigid -- contribution to the total cosmological constant. Still, the action is perfectly local and additive, and the diffeomorphisms remain unbroken. Further, the variation with respect to $\Lambda(x)$ yields $F_{\mu\nu\lambda\sigma} = \sqrt{g} \epsilon_{\mu\nu\lambda\sigma}$, meaning that $F$ is a non-propagating, auxiliary field\footnote{Since the action is linear in $F$, it means that integrating over it yields only a constraint, without any local degrees of freedom.}. We stress that the real reason for the absence of any local degrees of freedom from $\Lambda$ is  the gauge symmetry of the 4-form. The 4-form is invariant 
under the transformations $A_{\mu\nu\lambda} \rightarrow A_{\mu\nu\lambda} + 3 \partial_{[\mu} B_{\nu\lambda]}$. If we integrate the last term in the gauge fixing action (\ref{gfterm}) we find 
\be
- \,\, \frac{1}{3!} \int \partial_{[\mu} \Lambda(x) \,  A_{\nu\lambda\sigma]} \, dx^\mu dx^\nu dx^\lambda dx^\sigma
\ee
and under a gauge transformation it changes by $\delta S_{GF} = -\frac12  \int \partial_{[\mu} \Lambda(x) \,  \partial_{\nu} B_{\lambda\sigma]} \, dx^\mu dx^\nu dx^\lambda dx^\sigma$. So gauge invariance $\delta S_{GF} = 0$ is really what forces $\partial_\mu \Lambda = 0$. Note, that nothing changes if we alter the gauge fixing condition for $\sqrt{g}$ to be dependent on $\Lambda$. Indeed we can take  
\be
S_{GF} = - \int  \, \left(\Lambda(x)  \sqrt{g} \, d^4x \, - \, \frac{1}{4!} \, \sigma\left(\frac{\Lambda(x)}{\mu^4}\right)  F_{\mu\nu\lambda\sigma} \, dx^\mu dx^\nu dx^\lambda dx^\sigma \right) \, .
\label{gftermgen}
\ee
without changing any of the results above.

The replacement of the $\Lambda$-dependent terms in (\ref{actionJ}) by the gauge-fixing action $S_{GF}$ of (\ref{gftermgen}) is precisely the 
trick we are after. It renders the action manifestly local, and forces $\Lambda = {\rm constant}$. Since we have one more rigid Lagrange multiplier in (\ref{actionJ}), $\kappa^2$ , we can follow exactly the same procedure to make it local off shell, and 
constant on shell: we add an extra copy of the 4-form term in (\ref{gftermgen}) to the action, where we replace $\Lambda/\mu^4$ by 
$\kappa^2/M_{Pl}^2$. So our manifestly local action that will sequester matter sector vacuum energy is 
\ba
S &=& \int d^4 x \sqrt{g} \left[ \frac{\kappa^2(x)}{2} R  - \Lambda(x) - {\cal L}_m( g^{\mu\nu} , \Phi) \right]  \nonumber \\
&& ~~~~~~~ + \, \frac{1}{4!} \, \int \, dx^\mu dx^\nu dx^\lambda dx^\sigma  \, \left[ \sigma\left(\frac{\Lambda(x)}{\mu^4}\right) {F_{\mu\nu\lambda\sigma}} +\hat \sigma\left(\frac{ \kappa^2(x)}{ M_{Pl}^2}\right) {\hat F_{\mu\nu\lambda\sigma}}  \right] 
 \, . ~~
\label{actionJloc}
\ea
Here, $F_{\mu\nu\lambda\sigma} = 4 \partial_{[\mu} A_{\nu\lambda\sigma]}$ and $\hat F_{\mu\nu\lambda\sigma} = 4 \partial_{[\mu} \hat A_{\nu\lambda\sigma]}$ are the two 4-forms whose gauge symmetries render $\Lambda$ and $\kappa^2$ constant on shell, respectively. The functions 
$\sigma$ and $\hat \sigma$ are two smooth functions, which are otherwise almost completely arbitrary\footnote{They mustn't be linear functions in order to permit solutions where the finite values of Newton's constant and background vacuum curvature are given by arbitrary form fluxes.}. Their form might be constrained by additional phenomenological arguments. The scales $\mu \la M_{Pl}$ are the field theory and gravitational cutoffs, respectively. Note that the 
$\kappa^2$-sector, which has dramatic consequences in our theory, is completely absent in unimodular gravity \cite{HT}. This makes the two theories very different. 

The field equations which follow from (\ref{actionJloc}) are now completely local: 
\ba
&&\kappa^2 G^\mu{}_\nu = (\nabla^\mu \nabla_\nu-\delta^\mu{}_\nu \nabla^2 )\kappa^2 + T^\mu{}_\nu-\Lambda(x) \delta^\mu{}_\nu  \nonumber \\
&& \frac{\sigma'}{\mu^4} \, F_{\mu\nu\lambda\sigma} =  \sqrt{g} \epsilon_{\mu\nu\lambda\sigma} \, , \qquad \, \,
  \frac{\hat \sigma'}{M_{Pl}^2} \, \hat F_{\mu\nu\lambda\sigma} =-\frac{1}{2 } R \sqrt{g} \epsilon_{\mu\nu\lambda\sigma} \, , \nonumber \\  
&& \frac{\sigma'}{\mu^4} \, \partial_\mu \Lambda=0 \, ,   \qquad \qquad \qquad \frac{\hat \sigma'}{M_{Pl}^2} \, \partial_\mu \kappa^2 = 0 \, . \label{eoms}
\ea
Here  $T_{\mu\nu}=\frac{2}{\sqrt{g}}\frac{\delta }{\delta g^{\mu\nu}}\int d^4 x \sqrt{g}{\cal L}_m( g^{\mu\nu} , \Phi)  $ is the matter stress-energy tensor.
The last two equations force $\Lambda$ and $\kappa^2$ to be integration constants, $\kappa^2$ is the bare Planck scale. To extract the relationship of the bare cosmological constant $\Lambda$ to the geometry and the matter fields we trace out the gravitational field equations, and average them over all of spacetime, 
using the equations for the 4-forms to eliminate $\langle R \rangle$. This yields 
\be
\Lambda=\frac14 \langle T^\alpha{}_\alpha \rangle +\Delta \Lambda \, , \qquad \qquad 
 \Delta \Lambda=\frac14 \kappa^2  \langle R \rangle=-\frac{\mu^4}{2} \frac{\kappa^2  \hat \sigma'}{M_{Pl}^2 \sigma'}\frac{\int \hat F_4}{\int F_4} \, . \label{DelLam2}
\ee
Substituting this into the gravitational field equations in (\ref{eoms}) gives
\be \label{geneeqs}
\kappa^2 G^\mu{}_\nu=T^\mu{}_\nu-\frac{1}{4} \delta^\mu{}_\nu \langle T^\alpha{}_\alpha \rangle-\Delta \Lambda \delta^\mu{}_\nu \, , 
\ee
with  $\Delta \Lambda$ given by equation (\ref{DelLam2}). The historic averages of \cite{KP1,KP2}, which are integrals of the relevant quantitives over the whole of spacetime, divided by the total spacetime volume (which needs to be treated with care when the spacetime volume is infinite!), are 
a consequence of `measuring' the renormalized vacuum energy on a solution, instead of arising from a nonlocal action.
Equation 10, along with the definition of $\Delta \Lambda$, describe the the full set of field equations (8), with the non-metric fields $\Lambda, \kappa$, and the 4-forms integrated out up to their global averages.

Now, as in the case of vacuum energy sequestering with global constraints, we work in the limit of (semi) classical gravity, treating 
gravitational fields only classically. To compute the quantum corrections involving loops with internal matter lines only, 
we use the equivalence principle to pick the largest locally flat frame on a fixed background geometry, and transform (\ref{actionJloc}) to
locally Minkowski coordinates. We compute the matter loops in the locally flat frame, using the standard techniques of QFT in flat space, while renormalizing the field theory at any required level in the loop expansion (possibly having to account for the hierarchies in the field theory directly by using specific tools like supersymmetry). Once the pure matter sector quantities are accounted for we turn to matter corrections to the gravitational background.

To begin with, we include the graviton vacuum diagrams which renormalize the Planck scale. On a background which solves 
(\ref{eoms})-(\ref{geneeqs}), the renormalized Planck scale is \cite{myers}  
\be
\left(M^{ren}_{Pl}{}\right)^2 \simeq \kappa^2 + {\cal O}(N) 
\left({M}_{UV}^{}\right)^2 + \sum_{species} {\cal O}(1)  
m_{}^2 \ln({M}_{UV}^{}/m_{}) + \sum_{species} {\cal O}(1)  
m_{}^2 + \ldots\, ,
\ee
where ${M}_{UV}^{} \simeq \mu$ is the matter UV regulator, $N$ counts the matter sector degrees of freedom,
and $m_{}$ a mass of a virtual particle in the loop. Since $\kappa^2$ is a classical integration constant, we could have initially taken it small, having to renormalize it by a large one-loop matter sector correction. However, as for any other UV-sensitive quantity, the physical value of $\left(M^{ren}_{Pl}\right)^2$ isn't calculable in QFT. It needs to be determined by measurement. Once it is set to its experimentally determined value, reflecting the observed hierarchy between the matter scales and the Planck scale, it is radiatively stable as long as 
$\mu \la M_{Pl}$, as we required. In effect, radiative stability follows since the Planck scale arises in response to the matter quantum corrections, as in induced gravity \cite{sakharov,adler}. Also note that as in \cite{KP1,KP2}, this additive renormalization of Planck scale does not change the field equations (\ref{eoms})-(\ref{geneeqs}) in the least.

Next we consider the loop corrections to the stress energy. We compute them in the locally flat frame, and account for both the local scales controlling the local stress energy contributions (such as energy densities and pressures of particles etc), and the constant vacuum energy. This gives rise to renormalized $\langle 0 | T^\mu{}_\nu | 0 \rangle$, which we split as $T^\mu{}_\nu = -V_{vac} \delta^\mu{}_\nu + \tau^\mu{}_\nu$ as in \cite{KP1,KP2}. Here $V_{vac} = \langle 0 | {\cal L}_m | 0 \rangle $ is the UV sensitive quantum vacuum energy calculated to any given precision; it is the external momentum-independent part of  $\langle 0 | T^\mu{}_\nu | 0 \rangle$. Clearly it cancels from $T^\mu{}_\nu-\frac{1}{4} \delta^\mu{}_\nu \langle T^\alpha{}_\alpha \rangle = \tau^\mu{}_\nu-\frac{1}{4} \delta^\mu{}_\nu \langle \tau^\alpha{}_\alpha \rangle$ in Eq. (\ref{geneeqs}). 

The remaining finite cosmological constant includes $\Delta \Lambda$ of (\ref{DelLam2}). It is a ratio of two 4-form fluxes and normalized derivatives of the functions $\hat \sigma, \sigma$. To compute the fluxes, we have to take the solutions for the 4-forms and integrate them over the whole of the locally Lorentzian box.
The fluxes are therefore purely geometric, infra-red quantities, being controlled by the size of the box and insensitive to the UV-cutoff.  
The only loop corrections to the flux-dependent terms come from the $\kappa^2$ and $\Lambda$ dependence of the prefactors. Since their variations depend on the dimensionless variables $\kappa^2/M_{Pl}^2$ and $\Lambda/\mu^4$ they are bounded by  ${\cal O}(1)$ for smooth $\hat \sigma, \sigma$. So all sources of the gravitational field are radiatively stable under the matter sector corrections\footnote{We are assuming that the UV regulator couples to the same metric $g_{\mu\nu}$ as the matter fields, as in \cite{KP1,KP2}.}. 
 
In fact it is instructive to split the field equations 
(\ref{eoms})-(\ref{geneeqs}) into separate sectors relative to how they depend on the space time metric. Since $\kappa^2$ and $\Lambda$ are constant on shell, we find, using form notation for the two 4-forms,
\ba
&&^\star F_4-\langle ^\star F_4 \rangle = 0 \, ,  ~ \quad \qquad  \, \langle ^\star F_4 \rangle = \frac{\mu^4}{\sigma'} \, ,  ~~~~~ \quad  ^\star \hat F_4-\langle ^\star \hat F_4 \rangle =  \frac{M_{Pl}^2}{2\kappa^2 \hat \sigma'} \left( \tau^\alpha{}_\alpha-\langle  \tau^\alpha{}_\alpha \rangle \right)\, , ~~~~~~~~
\label{avlessF} \\ 
&&\Delta \Lambda = \frac14 \kappa^2 \langle R \rangle = -\frac{\kappa^2 \hat \sigma' }{2 M_{Pl}^2} \langle ^\star \hat F_4 \rangle  \, ,  \quad  \quad ~~ \, 
\qquad  ~~\, 4 \Lambda + 4V_{vac} =  \langle  \tau^\alpha{}_\alpha \rangle + \kappa^2 \langle R \rangle \, , \label{avtrace} \\
&&\kappa^2 \left(R^\mu{}_\nu-\frac14 \, R \, \delta ^\mu{}_\nu\right)= \tau^\mu{}_\nu-\frac14 \tau^\alpha{}_\alpha \,  \delta ^\mu{}_\nu \, , \quad \quad \,
\kappa^2 \left( R-\langle R \rangle \right) =  - \left( \tau^\alpha{}_\alpha-\langle  \tau^\alpha{}_\alpha \rangle \right) \, ,  \label{graveqs} 
\ea
where $^\star$ denotes the Hodge dual of a form. The form field equations (\ref{avlessF}) show that the form sectors are radiatively stable. This follows from how the form fluxes are computed above, in the locally Lorentzian boxes, and because matter-generated vacuum energy explicitly cancels from the stress energy sources in (\ref{avlessF}). The loops do correct these expressions due to the explicit dependence on $\kappa^2,\Lambda$ but the corrections are at most ${\cal O}(1)$ because they are suppressed by $M_{Pl}^2, \mu^4$ respectively.
As a result the first of Eqs (\ref{avtrace}) similarly shows that $\Delta \Lambda$ and $\langle R \rangle$ are also radiatively stable, being proportional to $\langle ^\star \hat F\rangle$. The second of Eqs (\ref{avtrace}) is the cosmological constant counterterm selection condition following from (\ref{actionJloc}). The terms on its RHS are radiatively stable, by the preceding discussion but
$V_{vac}$ clearly is not. This means that the large radiative corrections are automatically cancelled by $\Lambda$: the dynamics picks a boundary condition which selects the bare counterterm $\Lambda$ that absorbs  radiative corrections to the vacuum energy\footnote{The spacetime volume also locally responds to radiative corrections by virtue of the 
second of Eqs. (\ref{eoms}). These corrections, by local rescaling, can be interpreted as local field theory renormalizations of dimensional quantities. However they are never larger than ${\cal O}(1)$, by virtue of our choice of $\sigma$ and $\mu$ as a QFT cutoff.}.
Finally the Eqs. (\ref{graveqs}) are the gravitational field equations split as a traceless and trace part for convenience of comparison. But as a consequences of (\ref{avlessF}), (\ref{avtrace}), the UV-sensitive part of vacuum energy is explicitly cancelled, unlike in GR (unimodular or not). This shows that except for the UV-sensitive contributions to the vacuum energy, the rest of the QFT gravitates just like in GR.

The residual correction $\Delta \Lambda$ is completely arbitrary, but radiatively stable once ${\kappa^2}$ is fixed to be $\sim \left(10^{18} {\rm GeV}\right)^2$. It is a part of the finite leftover cosmological constant after renormalization, together with $\frac14 \langle \tau^\alpha{}_\alpha  \rangle$. This residual finite cosmological constant is a priori completely arbitrary, and must be fitted to observations. In our universe, it is $\sim 10^{-12} \, {\rm eV}^4$. $\Delta \Lambda$ could be positive, yielding a universe which expands forever and has an infinite worldvolume. 
This is now natural: the matter sector masses are fixed and finite. From Eqs. (\ref{eoms})-(\ref{geneeqs}) and the action, even if $\Delta \Lambda \ne 0$, the Planck scale $\kappa^2$ is finite independently of it. Being UV-sensitive, it is also fixed by observation, and once measured to be $\sim M_{Pl}$, it is radiatively stable, just as is $\Delta \Lambda$. Although the flux integrals $\int F_4$, $\int \hat F_4$ formally diverge in an infinite spacetime, they do so at the same rate: $F_4$ is constant, and $\hat F_4$ is dominated by an asymptotic value of the curvature scalar $R$ which is bounded in a universe that expands forever when matter satisfies the null energy condition. Since both fluxes diverge as the worldvolume, their ratio is bounded, and $\Delta \Lambda$ can be finite, small and UV-stable, in an infinite universe with nonzero matter sector scales. In such cases, $\Delta \Lambda$ is the only nonzero contribution to the effective cosmological constant probed by geometry. The historic average of non-constant stress energy $\langle \tau^\alpha{}_\alpha \rangle$ is zero as long as the matter sources obey the null energy condition. The historic integrals are dominated by the regions near the turning point. These are never attained in infinite universes. But by continuity the largest contributions come from the regions with the largest volume. There both $\tau^\alpha{}_\alpha$ and the historic averages vanish. 

How does such a mechanism evade Weinberg's no-go? After all, the theory appears to reduce to GR in the gravitational sector. 
The key operational ingredient of the mechanism is that the action (\ref{actionJloc}) involves different local measures, using 4-forms which do not depend on the metric off shell. This means that the terms which multiply such measures in the action 1) are constant, by virtue of the extra gauge symmetries of the 4-form, 2) do not appear in the trace of the gravitational field equations, because this term comes precisely from  the variation with respect to the metric determinant. Hence the theory has an almost completely non-gravitating sector, affecting only the trace of the gravitational field equations in the extreme infra-red. It is therefore {\it strongly violating} the weak equivalence principle in the far infrared, modifying only how the vacuum energy gravitates. The form fields play the role of a selection mechanism and the vacuum energy sink, adjusting order by order in the loop expansion to arrange for the bare counterterm 
$\Lambda$ to absorb away {\it only} the vacuum energy loops from the gravitational field equations\footnote{Note, that the approximate symmetries mentioned in the footnote 2 are still operational, ensuring that the residual cosmological constant $\Delta \Lambda$ is naturally small.}. Because of this, even though the local solutions of the theory are the same as in GR, the global structure which controls the vacuum geometry is very different. We emphasize that although the numerical value of the effective cosmological constant is a priori undetermined, once matched to observations it remains unaffected even as we include additional radiative corrections to the vacuum energy {\it without} the need of fine tuning by hand.  This is certainly unlike in plain vanilla GR, and is a consequence of our modification of the global dynamics of gravity.

In summary, we have constructed a manifestly local theory which sequesters all matter-generated quantum vacuum energy. The theory arises from a local action which is additive in spacetime, and admits standard Hamiltonian dynamics, being a consistent starting point for a definition of the Feynman path integral. On shell, by virtue of the local conservation laws, the modifications of the gravitational sector behave very similarly to the global setup of \cite{KP1,KP2}. However now solutions can have a finite, eternal cosmological constant, and an infinite worldvolume, while supporting finite Planck scale and matter sector scales. It is interesting to explore the framework further, to understand cosmological behavior, inflation (including eternal inflation \cite{eternal} which is consistent with this framework), effects of phase transitions and interplay between gravity and particle physics.

\vskip.3cm

{\bf Acknowledgments}: 
We wish to thank to G. D'Amico, S. Dimopoulos, S. Dubovsky, M. Kleban, A. Lawrence, A. Waldron and A. Westphal for useful discussions.
NK thanks the School of Physics and Astronomy, U. of Nottingham for hospitality in the course of this work. 
AP thanks Department of Physics, UC Davis for reciprocated hospitality. 
NK and GZ are supported by the DOE Grant DE-SC0009999. 
AP was funded by a Royal Society URF, and DS by an STFC studentship.

\end{document}